# Auger Recombination Coefficients in Type-I Mid-Infrared InGaAsSb Quantum Well Lasers

Timothy D. Eales[1], Igor P. Marko[1], Alfred R. Adams[1], Jerry R. Meyer[2], Igor Vurgaftman[2], Stephen J. Sweeney[1*]

[1]Department of Physics and Advanced Technology Institute, University of Surrey, Guildford, Surrey, GU2 7XH, UK

[2]Code 5613, Naval Research Laboratory, Washington DC 20375, USA

**Abstract**

From a systematic study of the threshold current density as a function of temperature and hydrostatic pressure, in conjunction with theoretical analysis of the gain and threshold carrier density, we have determined the wavelength dependence of the Auger recombination coefficients in InGaAsSb/GaSb quantum well lasers emitting in the 1.7-3.2 µm wavelength range. From hydrostatic pressure measurements, the non-radiative component of threshold currents for individual lasers was determined continuously as a function of wavelength. The results are analysed to determine the Auger coefficients quantitatively. This procedure involves calculating the threshold carrier density based on device properties, optical losses, and estimated Auger contribution to the total threshold current density. We observe a minimum in the Auger rate around 2.1 µm. A strong increase with decreasing mid-infrared wavelength (< 2 µm) indicates the prominent role of intervalence Auger transitions to the split-off hole band (CHSH process). Above 2 µm, the increase with wavelength is approximately exponential due to CHCC or CHLH Auger recombination, limiting long wavelength operation. The observed dependence is consistent with that derived by analysing literature values of lasing thresholds for type-I InGaAsSb quantum well diodes. Over the wavelength range considered, the Auger coefficient varies from a minimum of $\lesssim 1\times10^{-16} \mathrm{cm}^4\mathrm{s}^{-1}$ at 2.1µm to $\sim 8\times10^{-16} \mathrm{cm}^4\mathrm{s}^{-1}$ at 3.2µm.

**Introduction**

Semiconductor lasers emitting in the midwave infrared (mid-IR) have become key components in numerous applications, including compact mid-IR absorption spectroscopy, free-space optical communications and military counter-measures. Several decades of performance improvements, resulting from better material quality and structural design, have established emitters with GaSb-based InGaAsSb type-I active quantum wells (QWs) as the semiconductor lasers of choice for the 2-3 µm spectral range.

One of the most important semiconductor laser figures of merit for such devices is the threshold current density, $J_{th}$. Fig. 1a plots $J_{th}$ for a selection of the best high-performance mid-IR type-I InGaAsSb QW lasers [1,2,11,12,3–10]. For lasing wavelengths between 2.05 and 2.65 µm, the threshold current densities at room temperature are below 100 A/cm², and they remain below 200 A/cm² in devices operating up to 3 µm. These are among the lowest thresholds reported for any edge emitting quantum well semiconductor lasers, regardless of wavelength [13].

---

[*] Corresponding author: s.sweeney@surrey.ac.uk

However, in spite of success in the development of mid-IR type-I diode lasers, their threshold current densities are sensitive to both temperature and emission wavelength. The temperature sensitivity is quantified by the characteristic temperature, $T_0$, which is defined as the temperature increment over which the threshold current density increases by a factor $e$ (Euler's number) [14]. Thus, a high $T_0$ indicates better temperature stability and a more gradual increase of the threshold current density with increasing temperature. For the same wavelength range as in Fig. 1a, characteristic temperatures near room temperature for some of the most stable type-I mid-IR diode lasers (highest $T_0$) are presented in Fig. 1b. Despite substantial differences in the device design parameters (number of quantum wells, band offsets, strain etc.), which are likely close to optimal in each case due to the high-performance characteristics, $T_0$ systematically decreases with increasing wavelength. Whereas $T_0$ typically exceeds 100 K at $\lambda \approx 2$ μm, it falls below 50 K at wavelengths beyond 3 μm. This compares with near-IR $T_0$ values exceeding 200 K for GaAs-based lasers operating around 1 μm [15–17]. Detailed consideration shows that a given device's characteristic temperature depends on various parameters and its heterostructure design, and $T_0$ itself can depend strongly on temperature due to the complex interplay of different recombination processes [18].

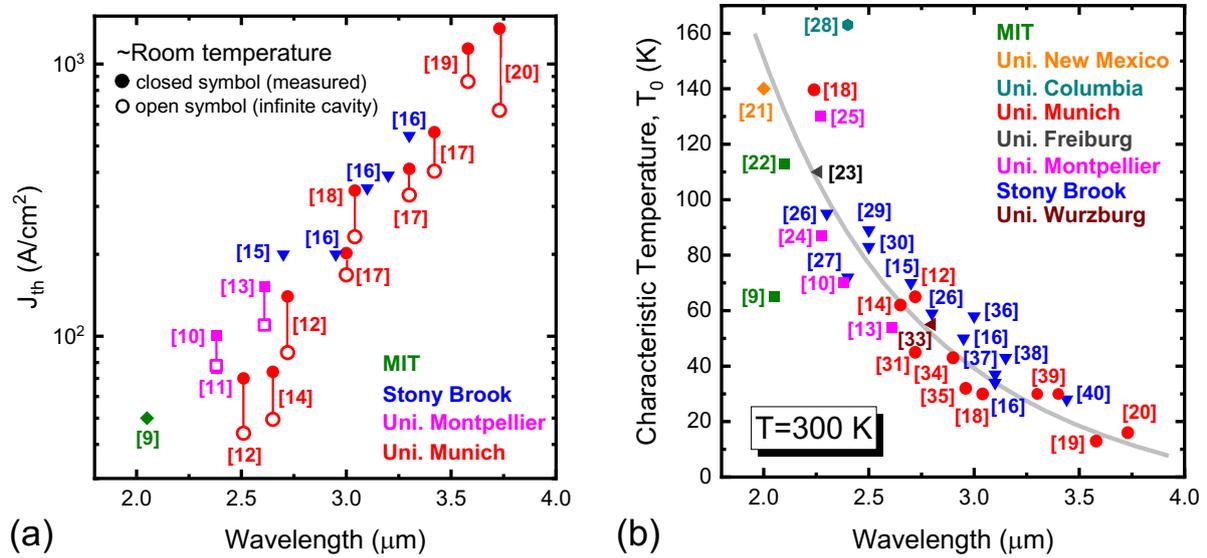

**Figure 1: (a)** *Threshold current densities at room temperature for GaSb-based type-I diode lasers operating in the 2-3.5 μm spectral range* [1,2,11,12,3–10]. *Solid markers indicate the measured $J_{th}$ values, while open markers correspond to $J_{th}$ extrapolated to an infinite cavity length.* **(b)** *Highest characteristic temperatures $T_0$ reported for type-I diode lasers emitting in the same wavelength range at temperatures near ambient* [1,2,19–28,4,29–38,5–8,10–12]. *Here the solid curve is a guide to the eye. Note that some of the variation is introduced by differences in the device geometry, operational conditions such as duty cycle, facet coatings, etc. It should also be remarked that lasers with the lowest thresholds do not necessarily display the highest $T_0$ values.*

Several factors limit the performance and restrict the spectral coverage of type-I mid-IR diode lasers. These are the band offsets, optical confinement factor, and optical loss mechanisms including free carrier absorption. The strong variation of $J_{th}$ with $\lambda$ in Fig. 1a indicates that the mechanism causing the performance to degrade at longer wavelengths must have an inherently strong dependence on the bandgap. In fact, it is well known that the threshold current densities of most interband mid-IR semiconductor lasers are dominated by the non-radiative multi-carrier Auger recombination mechanism [39,40,18]. In this process, the recombination of an electron and a hole is accompanied by the transition of a third carrier (either electron or hole) to an excited state. Since both energy and momentum must be conserved in the Auger process, the resulting rate, as quantified by the Auger coefficient, is quite sensitive to the details of the band structure. Furthermore, the Auger rate tends to increase rapidly with both temperature and wavelength (decreasing bandgap), often exponentially [39–41]. Despite the Auger coefficient's dominant effect on key diode laser properties, it has never been widely characterized previously for type-I QWs with mid-IR bandgaps [42], and reported values have varied considerably [42]. The substantial uncertainty in this parameter has made it difficult to reliably predict threshold current densities and other type-I mid-IR diode laser properties such as $T_0$. We note, however, that Auger coefficients for type-II mid-IR lasers emitting at wavelengths beyond 3 μm have been characterized more extensively [43,44].

**Methods**

*Auger Coefficient Calculations*

Since the Auger process involves three carriers, at non-degenerate concentrations of the sheet carrier density per active quantum well ($n$) the Auger recombination rate is proportional to $n^3$ and the corresponding lifetime may be written $\tau_A \approx 1/Cn^2$, where $C$ is the 2D *Auger recombination coefficient* (with units cm$^4$/s, as opposed to the 3D Auger coefficient for bulk materials that has units cm$^6$/s). The total threshold current density is the sum of the radiative and non-radiative contributions, where the latter is strongly dominated by Auger recombination rather than Shockley-Read recombination in high-quality mid-IR materials [45]. The non-radiative component is then $J_{Auger} = eN_{QW}Cn_{th}^3$, where $e$ is the electronic charge, $N_{QW}$ is the number of active quantum wells (assumed to be populated equally), and $n_{th}$ is the threshold carrier density. By combining the calculated threshold carrier density with the measured threshold current density of a given laser, we can extract the Auger coefficient for that laser's gain material. The optical gain is calculated using the 8-band **k·p** theory solved using the reciprocal space method [46]. Assuming equal electron and hole densities, $n_{th}$ is determined using the experimental cavity loss, where available, in conjunction with the confinement factor calculated for the layer structure and laser waveguide specified for each device. The contribution of radiative recombination to the threshold current was determined experimentally by measuring the temperature dependence of the spontaneous emission at threshold, which is observed through a window milled in the laser substrate. If at low temperature the threshold current density is dominated by radiative recombination, then the integrated spontaneous emission at threshold can be normalised at low temperatures under the assumption $J_{th}=J_{rad}$ to estimate the absolute value and fraction of the radiative component at room temperature [45,47]. From such measurements on a range of mid-IR devices [18,39,43–45], we find that at ambient temperature the radiative component accounts for approximately 20% of $J_{th}$, which implies that non-radiative recombination dominated by the Auger component ($J_{Auger}$) accounts for the remaining 80%. We can then calculate the Auger coefficient $C$ by dividing $J_{Auger}$ by $n_{th}^3$, which is assumed to be insensitive to pressure. Fig 2 presents the results of this procedure, as applied to type-I mid-infrared devices from this work and Refs. [3], [9], [18], [19], [27], [36], [37], and [40].

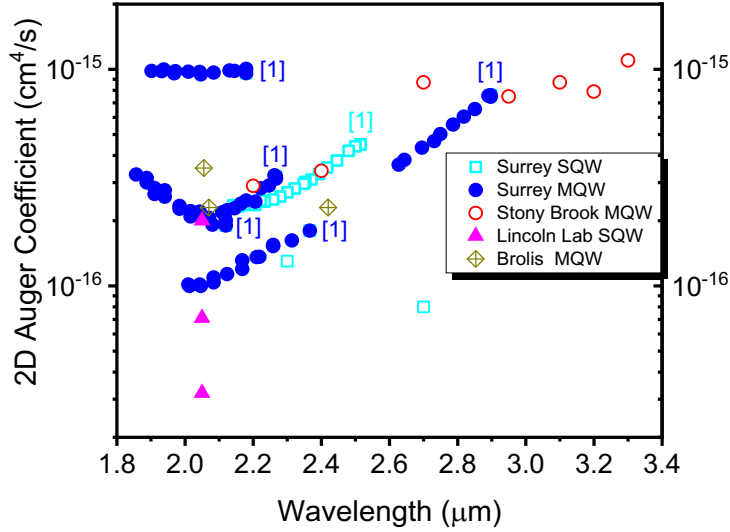

**Figure 2:** *Extracted Auger coefficients at threshold as obtained from $J_{Auger} \propto C n_{th}^3$. The threshold carrier density for each device is calculated using the reported structural details and optical loss value. Threshold current densities were reported for three different cavity lengths of the Lincoln Lab devices, which produce three different values of the Auger coefficient. This may indicate a deviation from the cubic dependence of the non-radiative component versus $n_{th}$.*

Figure 2 shows considerable scatter in the Auger coefficients extracted for the various devices. Some of this may be attributed to uncertainties and measurement errors in the optical losses and confinement factors that are employed, as well as deviations of the grown structures from the nominal designs. However, some may also be associated with real variations in the QW layering and alloy compositions, which cause the band alignments and separations to differ among lasers emitting at the same wavelength. A further factor may be the simple $Cn_{th}^3$ model used to describe the Auger current, which was derived assuming non-degenerate populations. We expect the rate of change of the Auger rate to decrease with increasing carrier density as the carriers start becoming degenerate (a necessary condition for population inversion). Modelling suggests that the Auger current's dependence on carrier density is not strictly cubic, and that the Auger coefficient itself depends on $n_{th}$. Thus the threshold condition particular to each laser can influence the extracted value.

Apart from the much higher Auger coefficient extracted for one anomalous device emitting in the 1.9-2.2 μm range (its wavelength was tuned by hydrostatic pressure, as will be discussed below), the data shown in Fig. 2 are largely consistent. From a minimum around 2.0-2.1 μm, $C$ increases gradually toward longer wavelengths and more rapidly at shorter wavelengths. At a given wavelength, the variations induced by different $N_{QW}$, quantum well widths, and compositions of the quantum wells and barriers appear to be no greater than approximately a factor of two. The exception is the anomalous device mentioned above, whose threshold current density is an order of magnitude higher than expected, even though it displayed no obvious deficiencies in layer design or growth/processing quality. Its main difference is a thinner active QW (8 vs. 10-12 nm). Finally, the Auger coefficients extracted from the Lincoln Lab devices [9] vary by up to a factor of six depending on the cavity length, as indicated by the error bars in Fig 2. This may be due in part to the non-cubic dependence of the Auger current density on carrier density, since this device with a single active QW requires a high, degenerate carrier population to produce sufficient gain, especially for the shortest cavity with the highest mirror loss.

*Hydrostatic Pressure Technique*

While a complete knowledge of both the loss and recombination mechanisms is required to properly account for the wavelength and temperature dependencies of the experimental threshold current density and characteristic temperature, hydrostatic pressure provides an additional tool that allows us to isolate the contributions of specific mechanisms such as Auger recombination.

Hydrostatic pressure compresses a sample equally in all three dimensions. This reversible process modifies the atomic spacing, whilst preserving the crystal symmetries. One effect of the reduced lattice spacing is that the direct bandgap increases. In this work, we measure the room-temperature pressure dependence of the threshold current density. The pressure measurements were performed using a helium gas compressor system (UniPress U11) [48]. Since the pressure coefficients are often very similar for alloy systems used in a particular laser structure, hydrostatic pressure can reversibly tune the laser's photon energy while preserving heterostructure aspects such as the band offsets. Thus, loss mechanisms such as heterointerface carrier leakage and defect-related recombination tend to be almost independent of pressure [49]. For hydrostatic pressures up to 10 kbar, the dependence of lasing energy on pressure is well characterised by a linear coefficient, typically in the range ~8 – 15 meV/kbar [49,50]. When hydrostatic pressure is applied, type-I mid-infrared InGaAsSb lasers can emit a continuum of wavelengths from ~1.7 μm to 3.2 μm.

From basic theory, the radiative component of the threshold current density scales approximately as $\propto E_g^2$ [50]. Using the experimentally-measured pressure dependence of $J_{th}$ and the absolute value of its radiative component extracted at ambient using spontaneous emission measurements (as discussed above), we can derive the wavelength dependence of both the radiative and non-radiative components of $J_{th}$ [39]. It should be noted, however, that the simple theoretical dependence described above assumes constant gain for a given quasi-Fermi level separation, and additionally that the threshold gain is unchanged. An alteration of the gain condition, for example through a bandgap-dependent loss or a change in the optical confinement factor, will modify the bandgap dependence. These effects were treated in detail elsewhere [51].

**Results & Discussion**

*Hydrostatic Pressure Measurements*

The solid points in Fig. 3 plot the extracted wavelength dependence of the non-radiative current density at room temperature for eleven different type-I mid-IR lasers measured under high hydrostatic pressure [18,39,49]. The non-radiative threshold current densities are normalised by setting the densities of neighbouring devices equal when they operate at the same pressure-induced wavelength. This overcomes differences due to variations in $J_{th}$ for the devices designed and fabricated by different growers, which had differing numbers of QWs, strain, band offsets, etc. Despite these differences, we generally observe good continuity in the bandgap dependence of the normalised non-radiative threshold current density. The wavelength dependence of the normalised non-radiative current density also reproduces fairly closely the literature values for total threshold current density that was plotted in Fig. 1a and is overlaid in Fig. 3 as the open blue circles. This points to an underlying physical process which fundamentally limits the performance of type-I mid-IR lasers, especially at wavelengths extending to 3 μm and beyond. As we noted for the Auger coefficients plotted in Fig. 2, the non-radiative component of the threshold current density in Fig. 3 shows two clear regimes, corresponding to increasing threshold current densities on either side of the minimum around 2 μm. The strong increase at shorter wavelengths is consistent with the onset of CHSH Auger processes, in which the recombination of a conduction-band electron and heavy hole excites a second heavy hole to the spin-orbit split off valence subband. We note that this process limits the operation of InP-based telecoms lasers in the 1.3-1.6 μm range [52].

This process can only conserve both momentum and energy when the fundamental bandgap approaches resonance with the spin-orbit splitting between the valence band maximum and split-off band. This condition in fact occurs in the type-I InGaAsSb QWs of interest when the bandgap wavelength is just below 2 μm. The increase at longer wavelengths is similarly attributable to CHCC (which involves excitation of an electron to a higher conduction-band state) or CHLH (involving a heavy hole excitation into a state in the light-hole subband) Auger processes.

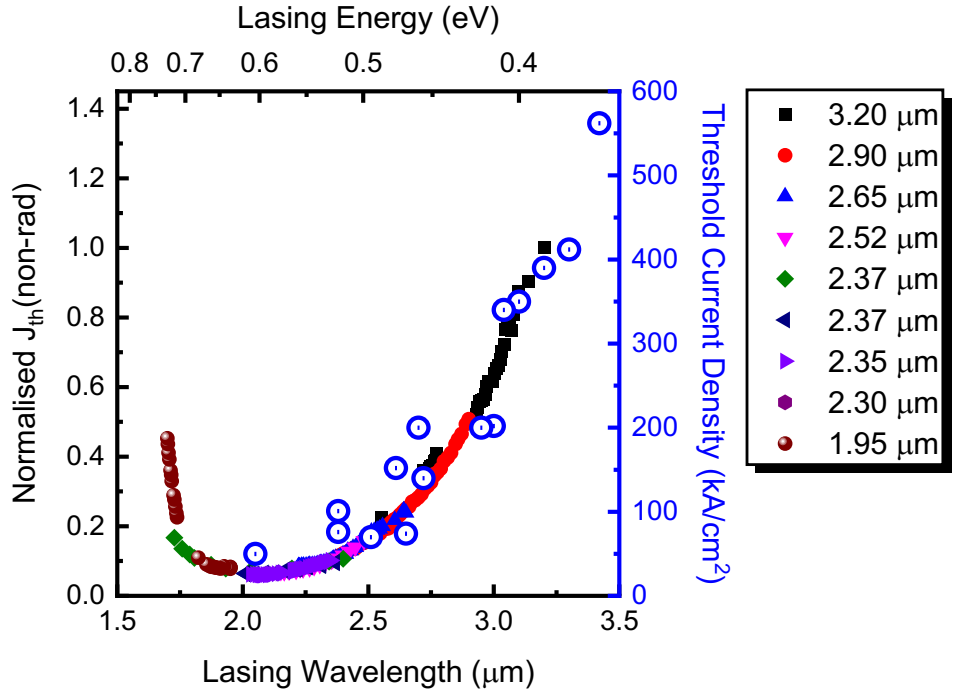

**Figure 3:** *Normalised non-radiative threshold current densities (solid points) as a function of lasing wavelength, as obtained from measurements of $J_{th}$ at high hydrostatic pressure, and using $J_{rad}$ at ambient as determined from the temperature dependences of $J_{th}$ and $J_{rad}$. The blue open circles are literature values of the total threshold current density taken from Fig. 1.*

For the purpose of extracting the Auger coefficient, it can be assumed as a first approximation that the non-radiative component is due entirely to Auger recombination. While this neglects any contributions from other non-radiative channels such as defect-related recombination, which may contribute to the non-radiative current [18,39], these are normally weak in high-quality lasers at the carrier-injection levels required to reach threshold [45]. In cases where other recombination processes comprise a non-negligible proportion of the low temperature threshold current, the radiative component should be reduced correspondingly at all temperatures. In our temperature-dependent analysis, we assumed a fully radiative threshold current ($J_{th}=J_{rad}$) at the lowest temperatures of 20-70 K.

The significance of carrier leakage (temperature-dependent heterointerface carrier leakage or thermal spill-over) to the room temperature threshold current density is contested in the literature [53]. However, the strong pressure dependence of the threshold current density argues against noticeable carrier leakage at room temperature in the devices considered here. The leakage current should have essentially flat pressure dependence because the band offsets do not change appreciably with pressure [54].

*Optical Losses*

The scatter of the Auger coefficients in Fig. 2 contrasts the trends derived in Fig. 3 from the hydrostatic and literature data. One potential reason, which we analyse in more detail below, could be uncertainty in the calculated threshold carrier density, which in turn depends on the optical loss. The value of $n_{th}$ is especially sensitive to loss when a single quantum well must produce all the gain. Furthermore, its cubic carrier density dependence amplifies the sensitivity of $J_{Auger}$ to $n_{th}$. A common method for evaluating the internal optical loss $\alpha_i$ is to measure the external differential efficiency as a function of inverse cavity length. However, the accuracy of that approach can be compromised by the assumption of pinned electron and hole quasi-Fermi-levels throughout the device, at and above threshold. For reasons that are not well understood, non-pinning of the spontaneous emission intensity above threshold has often been observed in both type-I [18,49] and type-II [55,56] mid-IR lasers, as well as in structures affected by inhomogeneity of the size or composition of the active region [49,57]. Therefore, as an alternative we have measured losses by the segmented contact method [58]. Using this approach, we measured internal losses of approximately 10 cm$^{-1}$ at wavelengths close to 2.5 μm, as shown in Fig. 4, which presents the net optical absorption ($A+\alpha_i$) spectrum at room temperature. The segmented contact devices were prepared using a focused ion beam technique to isolate the contact segments in a Fabry-Perot laser chip, which is processed with angled facets to prevent round trips of the emission. This loss is higher than the values typically measured by the inverse cavity length method, but is consistent with those measured using alternative optical gain methods such as the Hakki-Paoli technique [35].

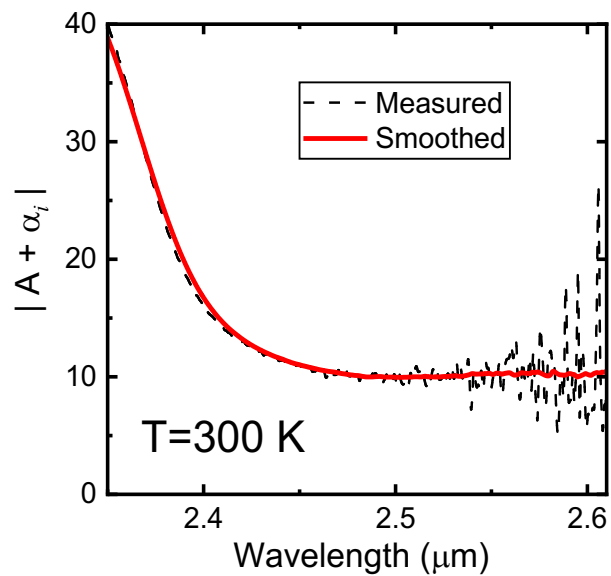

**Figure 4:** *Net optical absorption ($A+\alpha_i$) spectrum measured by the segmented contact method for a laser emitting at $\lambda \approx 2$ μm. The observed value of $\approx 10$ cm$^{-1}$ is higher than the losses typically reported using the inverse cavity length method.*

*Auger Coefficients*

Fig. 5 plots the Auger coefficient as a function of wavelength as obtained from analyses that rely on optical losses derived from absorption/gain measurement techniques. While only two values are presented (open circles), their wavelength dependence is consistent with the trend of the non-radiative current densities obtained from the hydrostatic pressure experiments. This confirms that in practice we expect the Auger coefficients for different InGaAsSb QW structures with sufficient carrier confinement to be within approximately a factor of 2 of the trend presented in Fig. 5.

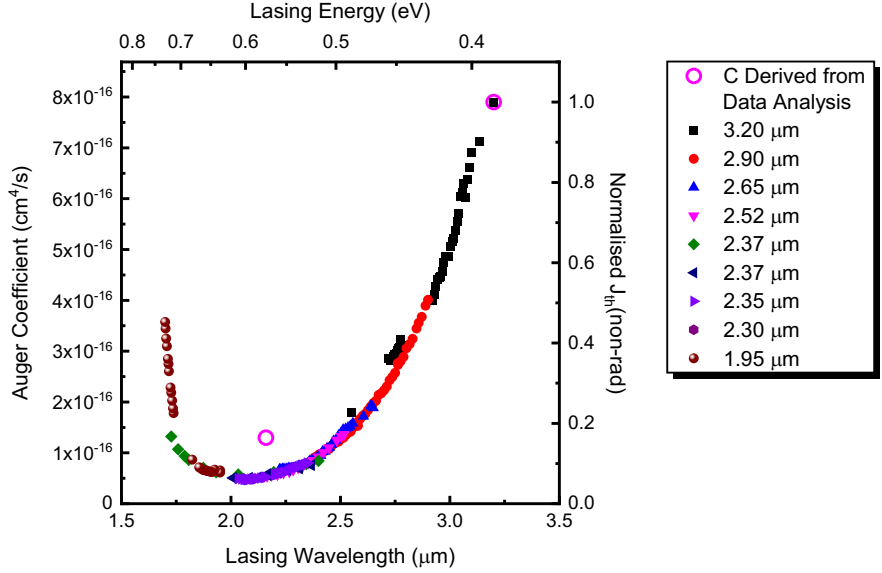

**Figure 5:** *Wavelength dependence of the non-radiative threshold current density, with normalisation with respect to two anchor points (open circles) corresponding to Auger coefficients obtained from data analysis with losses derived from optical gain techniques.*

**Conclusions**

From a combination of experimental temperature and hydrostatic pressure techniques, in conjunction with calculated threshold current densities, we have extracted wavelength-dependent Auger coefficients for GaSb-based mid-IR diode lasers with type-I InGaAsSb active quantum wells. The hydrostatic pressure measurements provide non-radiative threshold current densities at ambient temperature over a continuous range of wavelengths spanning 1.7 to 3.2 μm. Quantitative Auger coefficients were then derived using calculated threshold carrier densities that account for the number of active QWs, optical loss, and waveguide properties of each laser. Because the threshold carrier density, and by extension the Auger current density, are sensitive to the internal optical loss, the segmented contacted method was used to measure the optical losses in selected structures. In agreement with the hydrostatic pressure measurements, these results indicate a strong wavelength dependence of the Auger coefficient, and provide the best estimate for its absolute magnitude in type-I lasers based on the InGaAsSb gain system. A rapid increase of the Auger coefficient at wavelengths just below 2 μm indicates the importance of CHSH processes involving transitions across the spin split-off gap, which moves into resonance with the fundamental bandgap in the type-I quantum wells. This process is known to dominate the behaviour of devices operating in the near-infrared at telecommunications wavelengths [48]. We find that the Auger coefficient goes through a minimum value of $\lesssim 1\times10^{-16}$ cm$^4$s$^{-1}$ around 2.1 μm. At longer wavelengths, the onset of additional Auger recombination mechanisms such as the CHCC or CHLH processes cause the Auger coefficient to strongly increase, reaching a value of ~$8\times10^{-16}$ cm$^4$s$^{-1}$ around 3.2 μm.

**Acknowledgements**

The authors at the University of Surrey gratefully acknowledge EPSRC for funding this work under grants EP/N021037/1, EP/H005587/1 and for funding a studentship for TDE. Work at NRL was supported by the Office of Naval Research.

**Ethics Statement**

The authors confirm that ethical approval was not required for this work.

**Figures**

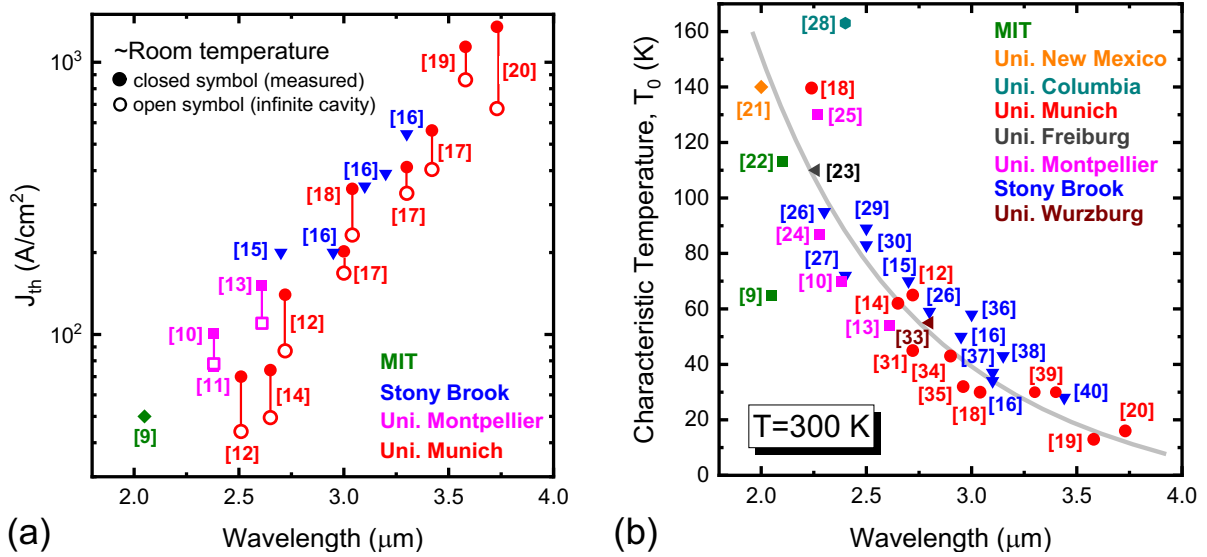

**Figure 1:**

**(a)** *Threshold current densities at room temperature for GaSb-based type-I diode lasers operating in the 2-3.5 μm spectral range* [1,2,11,12,3–10]. *Solid markers indicate the measured $J_{th}$ values, while open markers correspond to $J_{th}$ extrapolated to an infinite cavity length.*

**(b)** *Highest characteristic temperatures $T_0$ reported for type-I diode lasers emitting in the same wavelength range at temperatures near ambient* [1,2,19–28,4,29–38,5–8,10–12]. *Here the solid curve is a guide to the eye. Note that some of the variation is introduced by differences in the device geometry, operational conditions such as duty cycle, facet coatings, etc. It should also be remarked that lasers with the lowest thresholds do not necessarily display the highest $T_0$ values.*

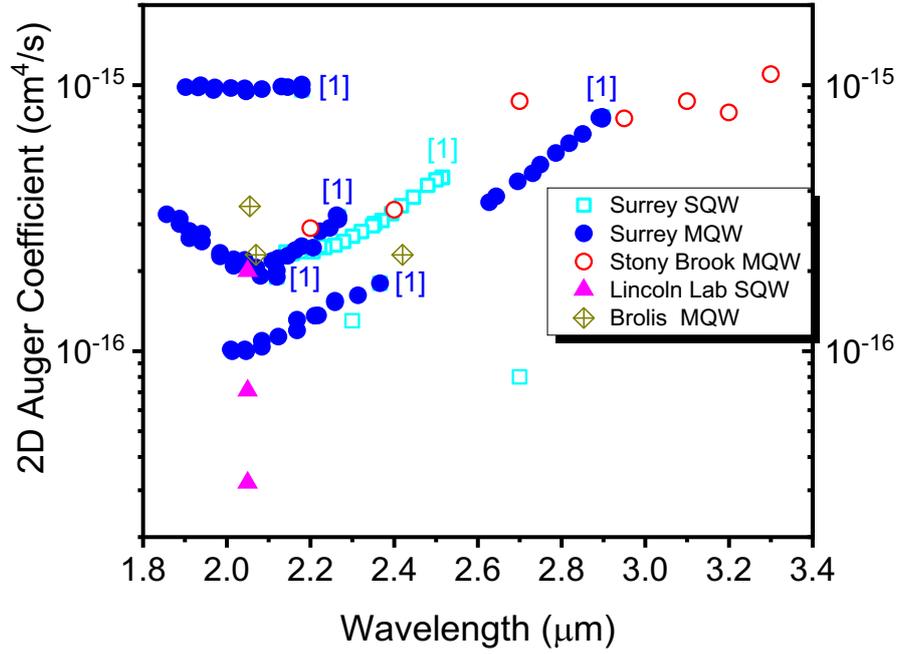

**Figure 2:**

*Extracted Auger coefficients at threshold as obtained from $J_{Auger} \propto Cn_{th}^3$. The threshold carrier density for each device is calculated using the reported structural details and optical loss value. Threshold current densities were reported for three different cavity lengths of the Lincoln Lab devices, which produce three different values of the Auger coefficient. This may indicate a deviation from the cubic dependence of the non-radiative component versus $n_{th}$.*

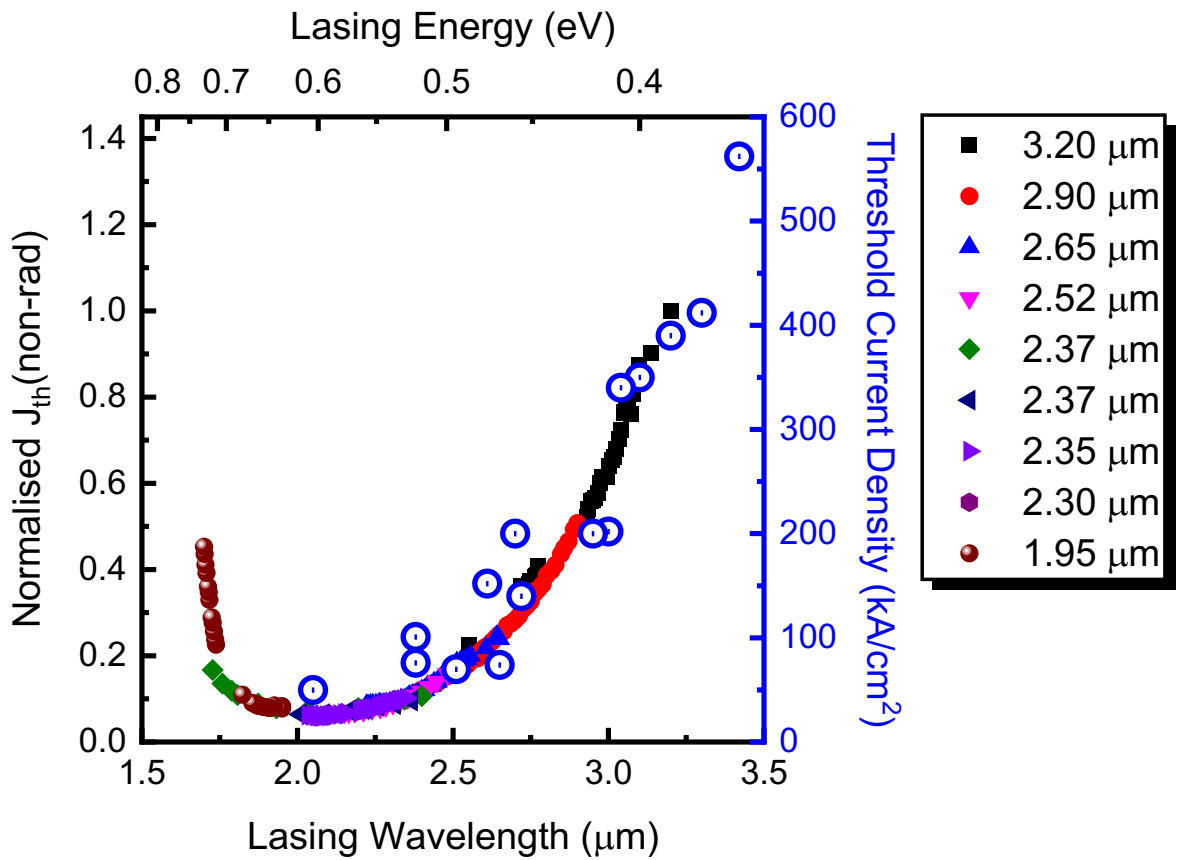

**Figure 3:**

*Normalised non-radiative threshold current densities (solid points) as a function of lasing wavelength, as obtained from measurements of $J_{th}$ at high hydrostatic pressure, and using $J_{rad}$ at ambient as determined from the temperature dependences of $J_{th}$ and $J_{rad}$. The blue open circles are literature values of the total threshold current density taken from Fig. 1.*

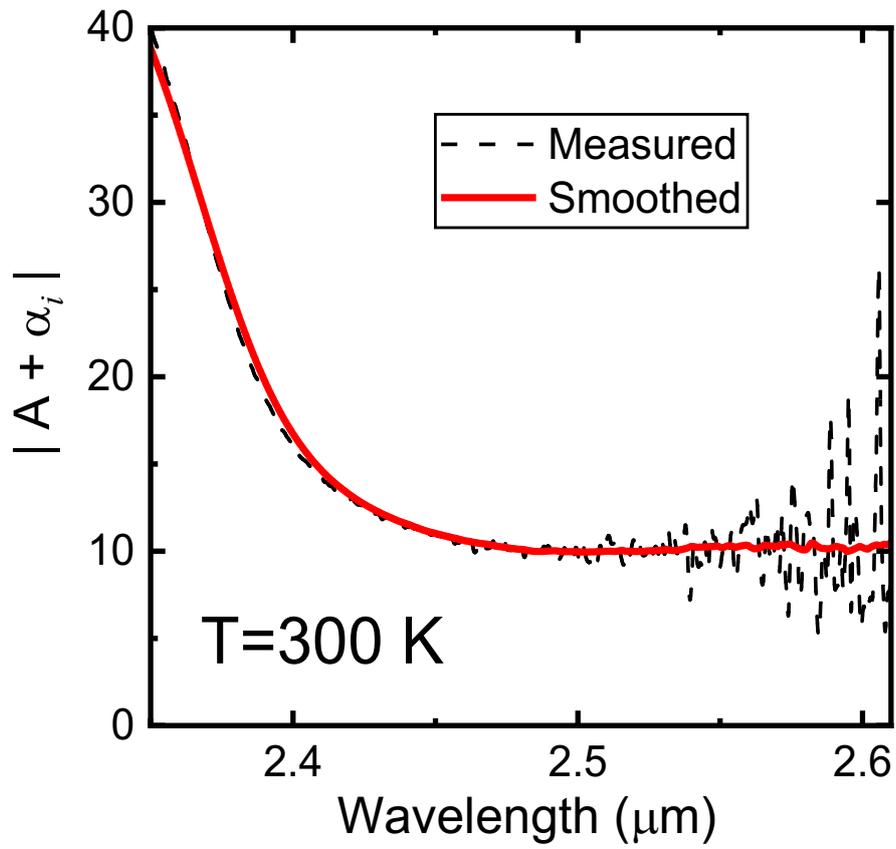

**Figure 4:**

*Net optical absorption (A+$\alpha_i$) spectrum measured by the segmented contact method for a laser emitting at λ ≈ 2 μm. The observed value of ≈ 10 cm$^{-1}$ is higher than the losses typically reported using the inverse cavity length method.*

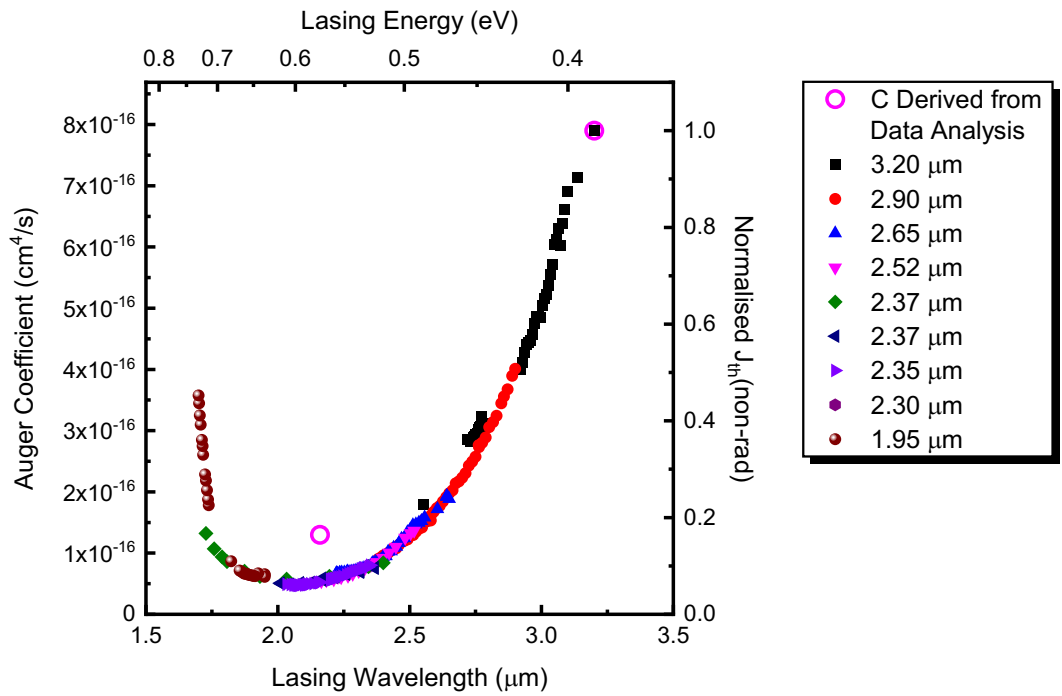

**Figure 5:**

*Wavelength dependence of the non-radiative threshold current density, with normalisation with respect to two anchor points (open circles) corresponding to Auger coefficients obtained from data analysis with losses derived from optical gain techniques.*